\documentclass{jfm}
\usepackage{subfigure}
\usepackage{graphicx}
\usepackage{natbib}
\usepackage{bm}% bold math

\ifCUPmtlplainloaded \else
  \checkfont{eurm10}
  \iffontfound
    \IfFileExists{upmath.sty}
      {\typeout{^^JFound AMS Euler Roman fonts on the system,
                   using the 'upmath' package.^^J}%
       \usepackage{upmath}}
      {\typeout{^^JFound AMS Euler Roman fonts on the system, but you
                   dont seem to have the}%
       \typeout{'upmath' package installed. JFM.cls can take advantage
                 of these fonts,^^Jif you use 'upmath' package.^^J}%
      }
  \else
  \fi
\fi

\ifCUPmtlplainloaded \else
  \checkfont{msam10}
  \iffontfound
    \IfFileExists{amssymb.sty}
      {\typeout{^^JFound AMS Symbol fonts on the system, using the
                'amssymb' package.^^J}%
       \usepackage{amssymb}%
         
         \let\geq=\geqslant
      }{}
  \fi
\fi

\ifCUPmtlplainloaded \else
  \IfFileExists{amsbsy.sty}
    {\typeout{^^JFound the 'amsbsy' package on the system, using it.^^J}%
     \usepackage{amsbsy}}
    {\providecommand\boldsymbol[1]{\mbox{\boldmath $##1$}}}
\fi

\newsavebox{\astrutbox}
\sbox{\astrutbox}{\rule[-5pt]{0pt}{20pt}}

\title[Aspect ratio dependence of heat transport by turbulent RB convection]{Aspect ratio dependence of heat transport by turbulent Rayleigh-B\'{e}nard convection in rectangular cells}
\author[Q. Zhou \emph{et al.}]{Quan ZHOU\thanks{Email address for correspondence: qzhou@shu.edu.cn}, Bo-Fang LIU, Chun-Mei LI and Bao-Chang ZHONG}
\affiliation{Shanghai Institute of Applied Mathematics and Mechanics, and Shanghai Key Laboratory of Mechanics in Energy Engineering, Shanghai University, Shanghai 200072, China}

\date{?? and in revised form ??}

\begin{document}

\maketitle

\begin{abstract}
We report high-precision measurements of the Nusselt number $Nu$ as a function of the Rayleigh number $Ra$ in water-filled rectangular Rayleigh-B\'{e}nard convection cells. The horizontal length $L$ and width $W$ of the cells are 50.0 cm and 15.0 cm, respectively, and the heights $H=49.9$, 25.0, 12.5, 6.9, 3.5, and 2.4 cm, corresponding to the aspect ratios $(\Gamma_x\equiv L/H,\Gamma_y\equiv W/H)=(1,0.3)$, $(2,0.6)$, $(4,1.2)$, $(7.3,2.2)$, $(14.3,4.3)$, and $(20.8,6.3)$. The measurements were carried out over the Rayleigh number range $6\times10^5\lesssim Ra\lesssim10^{11}$ and the Prandtl number range $5.2\lesssim Pr\lesssim7$. Our results show that for rectangular geometry turbulent heat transport is independent of the cells' aspect ratios and hence is insensitive to the nature and structures of the large-scale mean flows of the system. This is slightly different from the observations in cylindrical cells where $Nu$ is found to be in general a decreasing function of $\Gamma$, at least for $\Gamma=1$ and larger. Such a difference is probably a manifestation of the finite plate conductivity effect. Corrections for the influence of the finite conductivity of the top and bottom plates are made to obtain the estimates of $Nu_{\infty}$ for plates with perfect conductivity. The local scaling exponents $\beta_l$ of $Nu_{\infty}\sim Ra^{\beta_l}$ are calculated and found to increase from 0.243 at $Ra\simeq9\times10^5$ to 0.327 at $Ra\simeq4\times10^{10}$.

\end{abstract}

\section{Introduction}

Convection is ubiquitous in nature and in our everyday life. It can be found in the stars and planets, in the Earth's mantle and outer core, in the oceans and atmosphere, as well as in heat transport and mass mixing in many engineering applications. The paradigmatic example for natural convection is the Rayleigh-B\'{e}nard (RB) convection of an enclosed fluid layer between the colder top and the warmer bottom plates (Ahlers, Grossmann $\&$ Lohse 2009\emph{b}; Lohse $\&$ Xia 2010). A key issue in the study of turbulent RB convection is to understand how heat is transported upwards across the fluid layer by convective flows. The global heat transport by convection is usually expressed in terms of the Nusselt number, namely,
\begin{equation}
Nu=\frac{QH}{\lambda_f\Delta},
\end{equation}
where $Q$ is the heat current density across a fluid layer of thermal conductivity $\lambda_f$ with height $H$ and with an applied temperature difference $\Delta$. The dynamics of the system is determined by the geometrical configuration of the convection cell and by two dimensionless control parameters, i.e. the Rayleigh number and the Prandtl number,
\begin{equation}
Ra=\frac{\alpha g\Delta H^3}{\nu\kappa} \mbox{\ \ and\ \ } Pr=\frac{\nu}{\kappa}.
\end{equation}
Here, $g$ is the acceleration due to gravity and $\alpha$, $\nu$, and $\kappa$ are the isobaric thermal expansion coefficient, the kinematic viscosity, and the thermal diffusivity of the working fluid, respectively. The cell geometry is usually described in terms of one or more aspect ratios, such as $\Gamma\equiv D/H$ for a cylindrical cell of inner diameter $D$ and $(\Gamma_x\equiv L/H,\Gamma_y\equiv W/H)$ for a rectangular cell of horizontal length $L$ and width $W$. The $Ra$- and $Pr$-dependence of $Nu$ for various working fluids and cell geometries have been studied, both experimentally and numerically, in great detail for many years (Castaing \emph{et al.} 1989; Kerr 1996; Chavanne \emph{et al.} 1997, 2001; Du $\&$ Tong 2000; Kerr $\&$ Herring 2000; Niemela \emph{et al.} 2000; Ahlers $\&$ Xu 2001; Roche \emph{et al.} 2002, 2005; Xia, Lam $\&$ Zhou 2002; Verzicco $\&$ Camussi 2003; Niemela $\&$ Sreenivasan 2003; Shishkina $\&$ Wagner 2007; Ahlers, Funfschilling $\&$ Bodenschatz 2009\emph{a}; Funfschilling, Bodenschatz $\&$ Ahlers 2009; Song $\&$ Tong 2010; Stevens, Verzicco $\&$ Lohse 2010; Stevens, Lohse $\&$ Verzicco 2011\emph{a}; Silano, Sreenivasan $\&$ Verzicco 2010; He \emph{et al.} 2012). In addition, various theoretical models have been advanced to predict the behavior of convective heat transport \cite[]{castaing1989jfm, ss1990pra, gl2000jfm, gl2001prl, gl2003jfm, gl2004pof, gl2011pof, dubrulle2001epjb, dubrulle2002epjb}. For more detailed elucidation of the problem, we refer interested readers to the recent review paper by \cite{agl2009rmp}. On the other hand, there are fewer measurements focusing on the aspect-ratio-dependence (Xu, Bajaj $\&$ Ahlers 2000; Fleischer $\&$ Goldstein 2002; Cheung 2004; Nikolaenko \emph{et al.} 2005; Funfschilling \emph{et al.} 2005; Sun \emph{et al.} 2005\emph{a}; Niemela $\&$ Sreenivasan 2006; Roche \emph{et al.} 2010), and those measurements were all performed in containers with a cylindrical geometry. The objective of the present experimental investigation is to fill this gap by making high-precision measurements of $Nu$ over a wide range of the aspect ratio in convection cells with a rectangular geometry.

A lateral sidewall is indispensable for any convection experiment in the laboratory. The interaction between the sidewall and fluids would change the velocity and temperature distributions in the cell, and in turn change the flow structures of the system. Indeed, previous experimental studies for both cylindrical (du Puits, Resagk $\&$ Thess 2007) and rectangular (Xia, Sun $\&$ Cheung 2008) cells have shown that with increasing the aspect ratio the large-scale circulation (LSC) departs from a single-roll structure and becomes a multi-roll pattern. Thus, the $\Gamma$-dependence of $Nu$ may reflect the influence of flow structures on heat transport characteristics via the influence on the boundary layers. Measurements in cylindrical samples using water as working fluid ($Pr\approx4.3$) revealed that for $\Gamma\lesssim6$ $Nu$ decreases, albeit only a few percent, with increasing $\Gamma$ \cite[]{funfschilling2005jfm, sun2005jfm}. This suggests that the global heat transport properties of the RB system are not very sensitive to the flow structures for such parameter ranges. However, the situation is very different for lower $Pr$ and for the two-dimensional (2D) case. Using 3D direct numerical simulation (DNS), Bailon-Cuba, Emran $\&$ Schumacher (2010) found for $Pr=0.7$ that the minimum of $Nu$ occurs at the aspect ratio where the LSC undergoes a transition from a single-roll to a double-roll structure and the variations in $Nu$ for different $\Gamma$ are significant and can yield $11.3\%$ for $Ra=10^8$. For 2D steady-state calculations, \cite{ching2006jot} obtained a power-law of $Nu\sim\Gamma^{-1}$ for $\Gamma\leqslant3$. In 2D numerical RB flow, van der Poel, Stevens $\&$ Lohse (2011) identified different turbulent states for both $Pr=0.7$ and 4.3, corresponding to different roll structures and associated with different overall heat transfers. Transitions among these states thus lead to jumps and sharp transitions in $Nu(\Gamma)$. By connecting the structures of $Nu(\Gamma)$ to the way the flow organizes itself in the sample, \cite{lohse2011pre} explained why the aspect-ratio dependence of $Nu$ is more pronounced for small $Pr$. Compared with the 3D cases, we note that the 2D simulations show larger variations in $Nu(\Gamma)$ for both $Pr=0.7$ and 4.3, which may be explained by the different flow structures formed in the 2D and 3D cases.

Different geometrical shapes represent different symmetries, and hence may lead to different features of the flow and heat transfer. This prompts us to study the $\Gamma$-dependence of $Nu$ in a non-cylindrical system. In the present study, we choose rectangle as the shape of the cells, which has been widely used in the past (Xia, Sun $\&$ Zhou 2003; Gasteuil \emph{et al.} 2007; Maystrenko, Resagk $\&$ Thess 2007; Zhou $\&$ Xia 2008, 2010\emph{a},\emph{b}; Zhou \emph{et al.} 2007, 2010). It was found that the convective flows in cells with such geometry share some similar dynamics as those in 2D RB system, such as the reversals of the LSC \cite[]{xgl2010prl}.

The remainder of this paper is organized as follows. We give detailed descriptions of the experimental apparatus and conditions in $\S$2. Experimental results are presented and analyzed in $\S$3, which is divided into two parts. In $\S$3.1, we compare the measured $Nu$ for different aspect ratios and with those obtained in cylindrical cells. In $\S$3.2 we consider the finite conductivity corrections of the top and bottom plates and estimate $Nu_{\infty}$ for perfectly conducting plates. The scaling behaviors of $Nu_{\infty}$ are also discussed in $\S$3.2. We summarize our findings and conclude in $\S$4.

\section{Experimental apparatus and methods}
\begin{figure}
\begin{center}
\resizebox{0.55\columnwidth}{!}{%
  \includegraphics{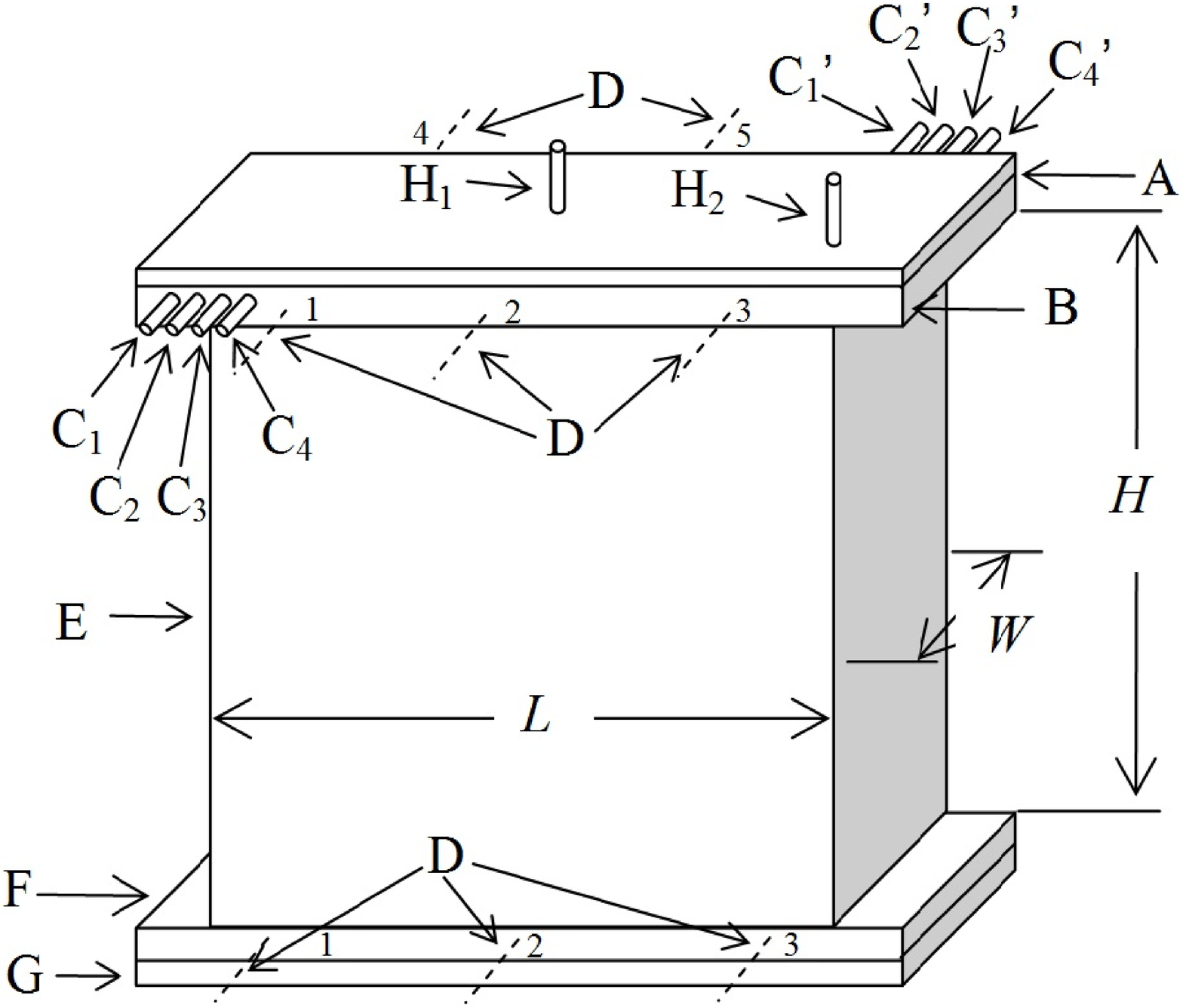}
}
\caption{Schematic diagram of front view of a rectangular cell. A: the Plexiglas cover; B: the copper top plate; C$_i$ and C$_i$' ($i=1,2,3,4$): nozzles connecting the channels to the refrigerated circulators; D: thermistors; E: the Plexiglas sidewall; F: the copper bottom plate; G: the copper cover for the bottom plate, H$_1$: nozzle for transferring fluid into the cell; H$_2$: nozzle for letting air out of the cell.} \label{fig:fig1}
\end{center}
\end{figure}

Figure \ref{fig:fig1} is a schematic drawing of the front view of our apparatus and the drawing corresponds to $(\Gamma_x,\Gamma_y)=(1,0.3)$. The sidewall of the cell, indicated as E in the figure, is composed by four transparent Plexiglas plates of 1.2 cm in thickness. The inner length $L$ and inner width $W$ of the cell are 50 cm and 15 cm, respectively. Six sidewalls of heights $H=49.9$, 25.0, 12.5, 6.9, 3.5, and 2.4 cm were used in the experiment. The corresponding aspect ratios are $(\Gamma_x\equiv L/H,\Gamma_y\equiv W/H)=(1,0.3)$, $(2,0.6)$, $(4,1.2)$, $(7.3,2.2)$, $(14.3,4.3)$, and $(20.8,6.3)$, respectively. As $\Gamma_y$ is proportional to $\Gamma_x$, in the remainder of the paper we use only $\Gamma_x$ to indicate the cell's aspect ratio for ease of presentation. The top (B) and bottom (F) plates are made of pure copper of 56 cm in length and 21 cm in width and their fluid-contact surfaces are electroplated with a thin layer of nickel to prevent the oxidation by water. The thickness of the top plate is 3.5 cm and that of the bottom plate is 1.5 cm. Silicon O-rings are placed between the copper plates and the sidewall plates to avoid fluid leakage. Eight stainless steel posts (not shown) hold the top and bottom plates together. They are insulated from the plates by Teflon sleeves and washers. Four parallel channels (not shown) of 1.5 cm in width and 2 cm in depth are machined into the top plate and the separation between adjacent channels is 1.5 cm. The channels start and end, respectively, at the two diagonal ends of the long edges. A silicon rubber sheet (not shown) and a Plexiglas plate (A) of 1.4 cm in thickness are fixed on the top to form the cover and also to prevent interflow between the adjacent channels. At two ends of the $i^{th}$ channel ($i=1,2,3,4$), there are two nozzles (C$_i$ and C$_i$') located, through which the channel is connected to a separate refrigerated circulator (Polyscience 9712) that has a temperature stability of 0.01 $^{\circ}$C. The channels and the circulators are connected such that the incoming cooler fluid and the outgoing warmer fluid in adjacent channels always flow in opposite directions. To provide constant and uniform heating, two rectangular Kapton film heaters of 25 cm in length and 15 cm in width are sandwiched between two copper plates (F and G) and are connected in parallel to a d.c. power supply (SGI 330X15D) with $99.99\%$ long-term stability. Therefore, the experiments were conducted under constant heating of the bottom plate while maintaining a constant temperature at the top plate. Note that recent high-resolution 2D \cite[]{doering2009prl} and 3D \cite[]{stevens2011jfm} simulations have revealed that turbulent thermal convection with boundary conditions of constant temperature and constant heat flux display identical heat transport at sufficient high Rayleigh numbers.

\begin{figure}
\begin{center}
\resizebox{0.495\columnwidth}{!}{%
  \includegraphics{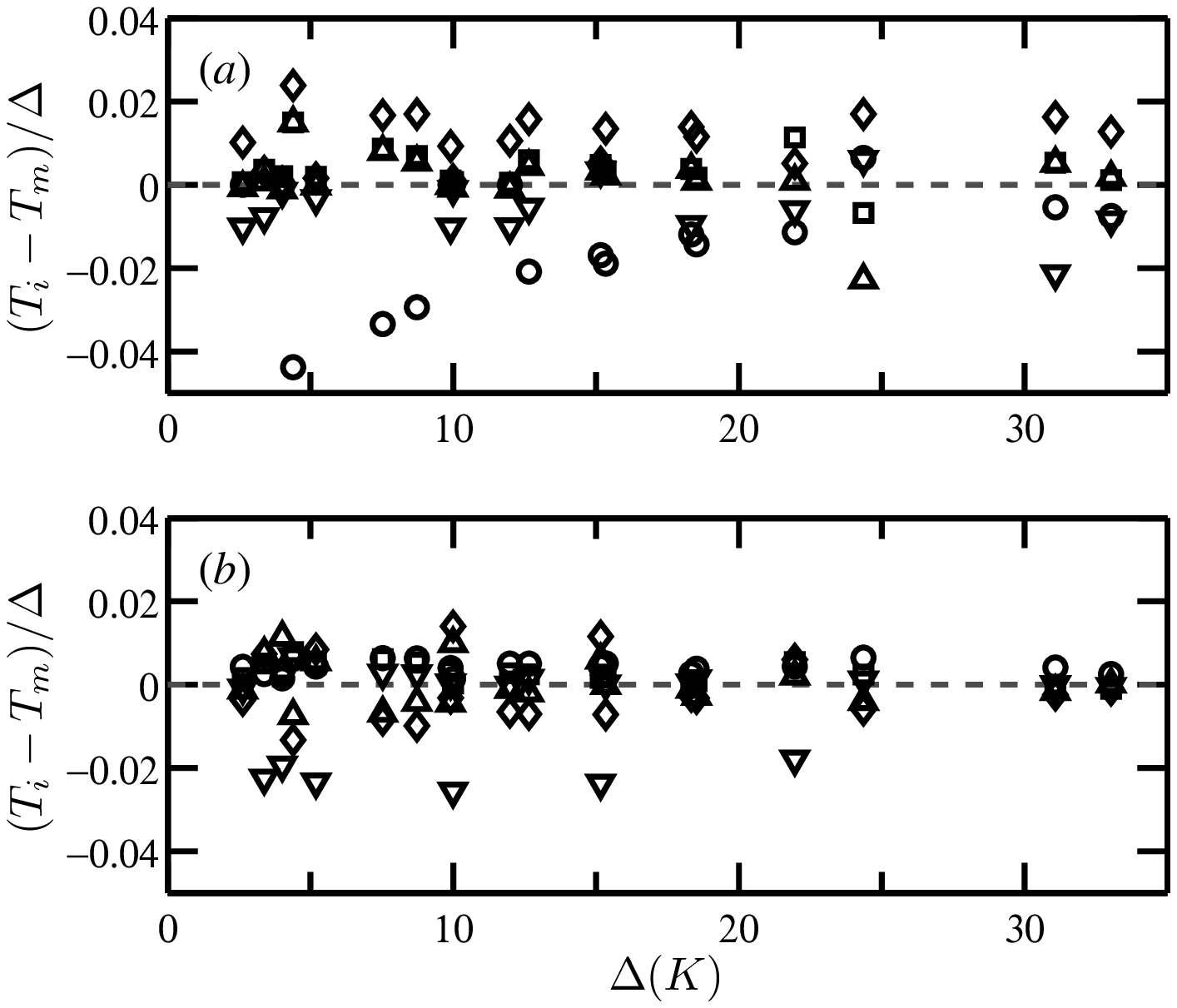}
}
\resizebox{0.495\columnwidth}{!}{%
  \includegraphics{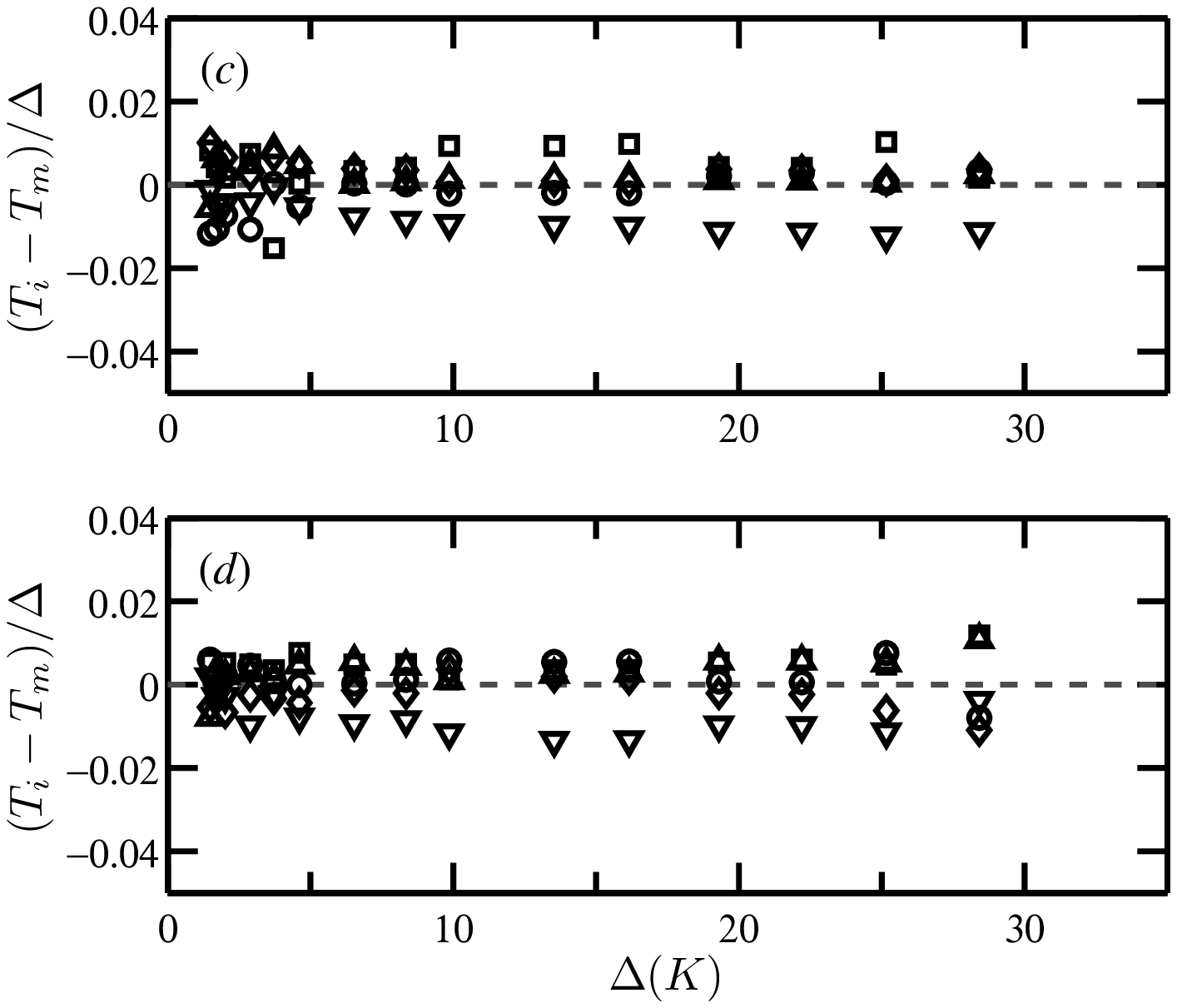}
}
\caption{The normalized horizontal temperature differences, $(T_i-T_m)/\Delta$, for the top (\emph{a}, \emph{c}) and bottom (\emph{b}, \emph{d}) plates and for $\Gamma_x=1$ (\emph{a}, \emph{b}) and 7.3 (\emph{c}, \emph{d}). The index $i$ ($=1,2,3,4,5$) of the thermistor is listed in figure \ref{fig:fig1} (see the top plate B and the bottom plate F).} \label{fig:fig2}
\end{center}
\end{figure}
% $i=1,2,3,4,5$ label the thermistors $\bigtriangledown$, $i=1$; $\bigcirc$, $i=2$; $\bigtriangleup$, $i=3$; $\square$, $i=4$; $\lozenge$, $i=5$.
Degassed water was used as the convecting fluid and the cell was leveled to better than $0.1^{\circ}$. During the measurements the entire cell was wrapped with several layers of Styrofoam. The temperature of each conducting plate was measured by five thermistors (D), which are embedded uniformly beneath the fluid-contact surface of the respective plate. When calculating the temperature difference $\Delta$ between the bottom and top plates, a correction was made for the temperature change between the thermistor position and the fluid-contact surface. In each measurement after $Ra$ was changed it took about 4 to 8 hours for the system to reach the steady state and we waited for at least 12 hours to start the measurements. A typical measurement lasted over 12 hours and more than 24 hours for low-$\Delta$ experiments ($\Delta<4$ $^{\circ}$C). No long-term drift of the mean temperature in the plates was observed over the duration of the measurement and the standard deviations were less than $0.5\%$ of $\Delta$ for all measurements.

To see the temperature uniformity of each conducting plate, we plot in figure \ref{fig:fig2} the normalized temperature variation, $(T_i-T_m)/\Delta$, for both the top and bottom plates. Here, $T_i$ ($i=1, 2, 3, 4, 5$) is the time-averaged temperature measured by the $i^{th}$ thermistor in a given plate and $T_{m}$ is the mean value of all the five thermistors in the same plate. It is found that the $\Gamma_x=1$ cell has the largest variation of plate temperature and we plot the $\Gamma_x=1$ results in figures \ref{fig:fig2}(\emph{a}) and (\emph{b}). The temperature variation for the other five $\Gamma_x$ are similar and we choose $\Gamma_x=7.3$ as an example and plot the results in figures \ref{fig:fig2}(\emph{c}) and (\emph{d}). One sees that $(T_i-T_m)/\Delta$ is at most $4\%$ for $\Gamma_x=1$ and is less than $2\%$ for all $\Gamma_x=7.3$ measurements. We thus estimate that a systematic error of the order of $1\%$ could be introduced into the measured $Nu$ by this horizontal temperature inhomogeneity of the conducting plates. Several thermistors were placed around the cell sidewall and around the bottom plate to monitor the environment temperature, based on which heat leakages to the environment and conduction by the posts and Plexiglas sidewall were calculated. The relative leakages to the total heat current were found to become more significant with decreasing $\Delta$ and thus we kept heat leakages from all sources to be less than $7\%$ of the total applied heat current by working with sufficiently large $\Delta$. We found that the largest source of leakage, especially for the small aspect-ratio cells, is through the cell sidewall which is insulated by multi-layers of Styrofoam. The errors in calculating the leaks come mainly from uncertainties in the thermal conductivities of the material involved, which are estimated to be less than $20\%$. This translates into an uncertainty of less than $1.5\%$ in the results of $Nu$.

\section{Results and discussion}
\subsection{$Nu$ vs. $Ra$ for different $\Gamma_x$}

\begin{figure}
\begin{center}
\resizebox{0.75\columnwidth}{!}{%
  \includegraphics{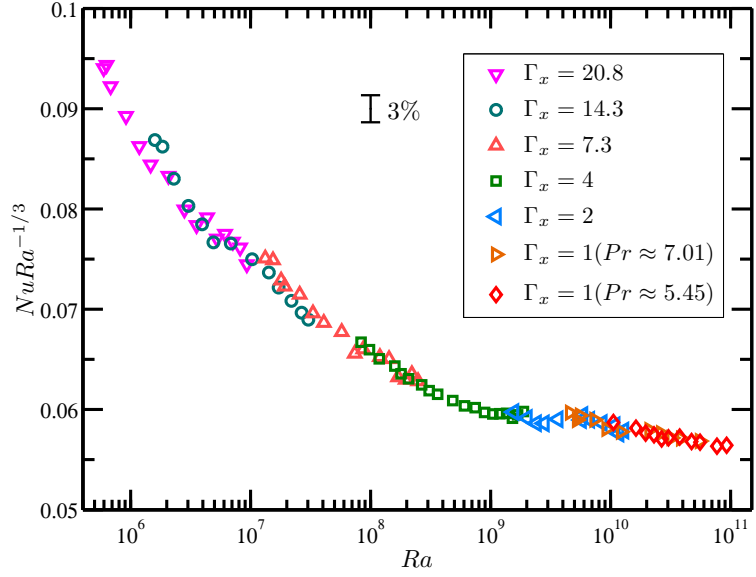}
}
\caption{(color online). Compensated $Nu/Ra^{1/3}$ on a linear scale vs. $Ra$ on a logarithmic scale. $\bigtriangledown$, $\Gamma_x=20.8$; $\bigcirc$, $\Gamma_x=14.3$; $\bigtriangleup$, $\Gamma_x=7.3$; $\square$, $\Gamma_x=4$; $\vartriangleleft$, $\Gamma_x=2$; $\vartriangleright$, $\Gamma_x=1$ and $Pr\approx7.01$; $\lozenge$, $\Gamma_x=1$ and $Pr\approx5.45$. Note that the data are as measured, without the correction for the finite plate conductivity.} \label{fig:fig3}
\end{center}
\end{figure}

\begin{table}
\centering \caption{Experimental parameters and results for the $\Gamma_x=1$ cell. Here, $Nu_{\infty}$ is calculated using Eq. (\ref{eq:1}) and the parameters $a=0.275$ and $b=0.39$ obtained by \cite{brown2005pof} in cylindrical cells of two sets of plates of different thermal conductivities (Cu and Al). It is also seen that $\alpha\Delta\lesssim0.01$ for all cases, which is considered to be sufficiently small. Note that two points are listed per line and the data are listed in chronological order.}
\label{tab:tab1}
\begin{tabular}{ccccccccccccc}
$\Delta$ (K) & $Ra$ & $Pr$ & $Nu$ & $Nu_{\infty}$ & $\alpha\Delta$ & $|$ & $\Delta$ (K) & $Ra$ & $Pr$ & $Nu$ & $Nu_{\infty}$ & $\alpha\Delta$ \\
&&& && ($10^{-3}$)& $|$ & &&& && ($10^{-3}$)\\
$\Gamma_x=1$&&& &&& $|$ & &&& &&\\
$3.44$ & $1.06\times10^{10}$ & $5.45$ & $128.8$ & $130.3$ & $1.04$ & $|$ & $5.24$ & $1.63\times10^{10}$ & $5.41$ & $147.4$ & $149.5$ & $1.59$ \\
$15.16$ & $4.73\times10^{10}$ & $5.40$ & $205.5$ & $210.6$ & $4.62$ & $|$ & $18.33$ & $5.57\times10^{10}$ & $5.48$ & $216.6$ & $222.4$ & $5.49$ \\
$24.36$ & $7.71\times10^{10}$ & $5.40$ & $239.8$ & $247.2$ & $7.42$ & $|$ & $9.97$ & $3.03\times10^{10}$ & $5.47$ & $178.3$ & $181.9$ & $2.99$ \\
$12.65$ & $3.77\times10^{10}$ & $5.52$ & $191.9$ & $196.2$ & $3.74$ & $|$ & $7.52$ & $2.30\times10^{10}$ & $5.46$ & $163.5$ & $166.4$ & $2.27$ \\
$6.41$ & $1.96\times10^{10}$ & $5.46$ & $155.4$ & $157.9$ & $1.93$ & $|$ & $8.74$ & $2.66\times10^{10}$ & $5.47$ & $170.4$ & $173.6$ & $2.63$ \\
$30.93$ & $9.27\times10^{10}$ & $5.51$ & $255.3$ & $264.0$ & $9.19$ & $|$ & $5.30$ & $9.25\times10^{9}$ & $7.01$ & $121.9$ & $123.1$ & $1.09$ \\
$7.02$ & $1.22\times10^{10}$ & $7.02$ & $133.1$ & $134.6$ & $1.44$ & $|$ & $3.20$ & $5.56\times10^{9}$ & $7.02$ & $105.3$ & $106.1$ & $0.66$ \\
$11.98$ & $2.11\times10^{10}$ & $6.98$ & $160.2$ & $162.8$ & $2.49$ & $|$ & $4.23$ & $7.34\times10^{9}$ & $7.03$ & $114.6$ & $115.6$ & $0.87$ \\
$15.10$ & $2.64\times10^{10}$ & $7.01$ & $171.6$ & $174.7$ & $3.12$ & $|$ & $2.63$ & $4.60\times10^{9}$ & $7.00$ & $99.3$ & $100.0$ & $0.54$ \\
$21.96$ & $3.87\times10^{10}$ & $6.99$ & $193.1$ & $197.3$ & $4.56$ & $|$ & $3.02$ & $5.26\times10^{9}$ & $7.02$ & $102.5$ & $103.2$ & $0.62$ \\
$31.40$ & $5.55\times10^{10}$ & $6.98$ & $216.9$ & $222.5$ & $6.54$ & $|$ & $18.52$ & $3.23\times10^{10}$ & $7.02$ & $182.0$ & $185.6$ & $3.82$ \\[3pt]

%\hline\hline
\end{tabular}
\end{table}

\begin{table}
\centering \caption{Experimental results for the $\Gamma_x=2$, 4, 7.3, 14.3, and 20.8 cells. Note that two points are listed per line and the data are listed in chronological order.}
\label{tab:tab2}
\begin{tabular}{ccccccccccccc}
$\Delta$ (K) & $Ra$ & $Pr$ & $Nu$ & $Nu_{\infty}$ & $\alpha\Delta$ & $|$ & $\Delta$ (K) & $Ra$ & $Pr$ & $Nu$ & $Nu_{\infty}$ & $\alpha\Delta$ \\
&&& && ($10^{-3}$)& $|$ & &&& && ($10^{-3}$)\\
$\Gamma_x=2$&&& &&& $|$ & &&& &&\\
$17.18$ & $6.67\times10^{9}$ & $5.26$ & $111.0$ & $114.1$ & $5.41$ & $|$ & $9.39$ & $3.64\times10^{9}$ & $5.26$ & $90.8$ & $92.6$ & $2.95$ \\
$21.63$ & $8.46\times10^{9}$ & $5.24$ & $119.5$ & $123.2$ & $6.84$ & $|$ & $7.12$ & $2.79\times10^{9}$ & $5.23$ & $82.5$ & $84.0$ & $2.25$ \\
$33.33$ & $1.28\times10^{10}$ & $5.28$ & $135.6$ & $140.5$ & $11.0$ & $|$ & $6.02$ & $2.40\times10^{9}$ & $5.19$ & $78.5$ & $79.7$ & $1.92$ \\
$29.86$ & $1.17\times10^{10}$ & $5.23$ & $130.7$ & $135.3$ & $9.45$ & $|$ & $14.62$ & $5.69\times10^{9}$ & $5.25$ & $105.3$ & $107.9$ & $4.61$ \\
$5.17$ & $2.01\times10^{9}$ & $5.25$ & $74.7$ & $75.8$ & $1.63$ & $|$ & $27.15$ & $1.08\times10^{10}$ & $5.18$ & $129.3$ & $133.8$ & $8.37$ \\
$14.94$ & $5.85\times10^{9}$ & $5.25$ & $106.1$ & $108.9$ & $4.57$ & $|$ & $3.82$ & $1.56\times10^{9}$ & $5.23$ & $69.2$ & $70.1$ & $1.21$ \\
$\Gamma_x=4$&&& &&& $|$ & &&& &&\\
$17.02$ & $8.90\times10^{8}$ & $5.25$ & $57.4$ & $59.1$ & $5.36$ & $|$ & $1.60$ & $8.33\times10^{7}$ & $5.25$ & $29.1$ & $29.4$ & $0.50$ \\
$35.83$ & $1.88\times10^{9}$ & $5.25$ & $73.8$ & $76.8$ & $11.3$ & $|$ & $2.26$ & $1.18\times10^{8}$ & $5.25$ & $31.9$ & $32.3$ & $0.71$ \\
$32.61$ & $1.68\times10^{9}$ & $5.29$ & $70.7$ & $73.4$ & $10.2$ & $|$ & $3.03$ & $1.59\times10^{8}$ & $5.24$ & $34.9$ & $35.3$ & $0.96$ \\
$29.21$ & $1.52\times10^{9}$ & $5.26$ & $68.0$ & $70.5$ & $9.18$ & $|$ & $5.10$ & $2.67\times10^{8}$ & $5.26$ & $40.2$ & $40.9$ & $1.61$ \\
$25.98$ & $1.36\times10^{9}$ & $5.25$ & $66.0$ & $68.4$ & $8.19$ & $|$ & $6.93$ & $3.62\times10^{8}$ & $5.25$ & $43.9$ & $44.7$ & $2.18$ \\
$22.96$ & $1.19\times10^{9}$ & $5.27$ & $63.2$ & $65.3$ & $7.20$ & $|$ & $9.16$ & $4.83\times10^{8}$ & $5.23$ & $47.8$ & $48.8$ & $2.90$ \\
$19.87$ & $1.04\times10^{9}$ & $5.24$ & $60.4$ & $62.3$ & $6.28$ & $|$ & $11.66$ & $6.06\times10^{8}$ & $5.27$ & $51.1$ & $52.3$ & $3.66$ \\
$14.25$ & $7.44\times10^{8}$ & $5.26$ & $54.5$ & $56.0$ & $4.49$ & $|$ & $3.96$ & $2.07\times10^{8}$ & $5.25$ & $37.3$ & $37.9$ & $1.25$ \\
$1.87$ & $9.78\times10^{7}$ & $5.26$ & $30.4$ & $30.7$ & $0.59$ & $|$ & $5.88$ & $3.08\times10^{8}$ & $5.25$ & $41.8$ & $42.5$ & $1.85$ \\
$3.44$ & $1.79\times10^{8}$ & $5.26$ & $35.9$ & $36.4$ & $1.08$ \\
$\Gamma_x=7.3$&&& &&& $|$ & &&& &&\\
$16.16$ & $1.42\times10^{8}$ & $5.24$ & $34.0$ & $35.1$ & $5.10$ & $|$ & $2.91$ & $2.57\times10^{7}$ & $5.25$ & $21.1$ & $21.4$ & $0.92$ \\
$13.53$ & $1.20\times10^{8}$ & $5.24$ & $32.1$ & $33.1$ & $4.27$ & $|$ & $4.63$ & $4.08\times10^{7}$ & $5.25$ & $23.6$ & $24.1$ & $1.46$ \\
$22.21$ & $1.96\times10^{8}$ & $5.25$ & $36.6$ & $37.9$ & $7.01$ & $|$ & $2.18$ & $1.92\times10^{7}$ & $5.25$ & $19.4$ & $19.6$ & $0.69$ \\
$9.87$ & $8.69\times10^{7}$ & $5.25$ & $29.2$ & $30.0$ & $3.11$ & $|$ & $1.75$ & $1.54\times10^{7}$ & $5.25$ & $18.6$ & $18.9$ & $0.55$ \\
$6.55$ & $5.74\times10^{7}$ & $5.26$ & $26.1$ & $26.7$ & $2.06$ & $|$ & $3.74$ & $3.31\times10^{7}$ & $5.24$ & $22.3$ & $22.7$ & $1.18$ \\
$1.51$ & $1.32\times10^{7}$ & $5.26$ & $17.8$ & $18.0$ & $0.47$ & $|$ & $25.14$ & $2.21\times10^{8}$ & $5.26$ & $38.4$ & $39.9$ & $7.91$ \\
$8.36$ & $7.37\times10^{7}$ & $5.25$ & $27.5$ & $28.2$ & $2.64$ & $|$ & $28.43$ & $2.49\times10^{8}$ & $5.27$ & $39.5$ & $41.1$ & $8.93$ \\
$2.04$ & $1.80\times10^{7}$ & $5.25$ & $19.1$ & $19.3$ & $0.64$ & $|$ & $19.31$ & $1.69\times10^{8}$ & $5.26$ & $35.0$ & $36.1$ & $6.07$ \\
$\Gamma_x=14.3$&&& &&&&\\
$12.35$ & $1.42\times10^{7}$ & $5.25$ & $17.8$ & $18.4$ & $3.89$ & $|$ & $2.65$ & $3.03\times10^{6}$ & $5.26$ & $11.6$ & $11.8$ & $0.83$ \\
$8.96$ & $1.03\times10^{7}$ & $5.26$ & $16.3$ & $16.8$ & $2.82$ & $|$ & $1.61$ & $1.84\times10^{6}$ & $5.26$ & $10.6$ & $10.7$ & $0.51$ \\
$5.98$ & $6.84\times10^{6}$ & $5.27$ & $14.5$ & $14.9$ & $1.88$ & $|$ & $14.95$ & $1.71\times10^{7}$ & $5.26$ & $18.6$ & $19.3$ & $4.70$ \\
$2.00$ & $2.29\times10^{6}$ & $5.26$ & $10.9$ & $11.1$ & $0.63$ & $|$ & $18.96$ & $2.19\times10^{7}$ & $5.24$ & $19.8$ & $20.6$ & $5.99$ \\
$3.44$ & $3.94\times10^{6}$ & $5.26$ & $12.4$ & $12.6$ & $1.08$ & $|$ & $23.45$ & $2.66\times10^{7}$ & $5.28$ & $20.8$ & $21.7$ & $7.34$ \\
$4.28$ & $4.90\times10^{6}$ & $5.26$ & $13.0$ & $13.3$ & $1.35$ & $|$ & $26.52$ & $3.02\times10^{7}$ & $5.27$ & $21.5$ & $22.4$ & $8.32$ \\
$1.39$ & $1.59\times10^{6}$ & $5.26$ & $10.1$ & $10.3$ & $0.44$ \\
$\Gamma_x=20.8$&&& &&&&\\
$11.75$ & $4.34\times10^{6}$ & $5.26$ & $12.9$ & $13.4$ & $3.70$ & $|$ & $3.19$ & $1.18\times10^{6}$ & $5.26$ & $9.1$ & $9.3$ & $1.01$ \\
$16.55$ & $6.12\times10^{6}$ & $5.26$ & $14.2$ & $14.8$ & $5.21$ & $|$ & $5.59$ & $2.07\times10^{6}$ & $5.26$ & $10.6$ & $10.9$ & $1.76$ \\
$19.15$ & $7.11\times10^{6}$ & $5.24$ & $14.8$ & $15.4$ & $6.05$ & $|$ & $1.85$ & $6.82\times10^{5}$ & $5.27$ & $8.1$ & $8.3$ & $0.58$ \\
$22.00$ & $8.18\times10^{6}$ & $5.24$ & $15.3$ & $16.0$ & $6.95$ & $|$ & $3.97$ & $1.47\times10^{6}$ & $5.26$ & $9.6$ & $9.8$ & $1.25$ \\
$14.18$ & $5.24\times10^{6}$ & $5.26$ & $13.4$ & $13.9$ & $4.46$ & $|$ & $2.48$ & $9.16\times10^{5}$ & $5.26$ & $8.7$ & $8.8$ & $0.78$ \\
$7.51$ & $2.81\times10^{6}$ & $5.23$ & $11.3$ & $11.6$ & $2.38$ & $|$ & $1.71$ & $6.30\times10^{5}$ & $5.26$ & $8.1$ & $8.2$ & $0.54$ \\
$9.59$ & $3.55\times10^{6}$ & $5.25$ & $12.0$ & $12.3$ & $3.02$ & $|$ & $25.00$ & $9.30\times10^{6}$ & $5.24$ & $15.7$ & $16.4$ & $7.90$ \\
$1.62$ & $5.96\times10^{5}$ & $5.26$ & $7.9$ & $8.1$ & $0.51$ \\
\end{tabular}
\end{table}

The measured $Nu$ with corresponding values of $\Delta$, $Ra$, and $Pr$ are given in tables \ref{tab:tab1} and \ref{tab:tab2}. The $\Gamma_x=1$ measurements were made at mean temperatures of 20$^{\circ}$C and 30$^{\circ}$C, corresponding to $Pr=7.01$ and 5.45, respectively, and the measurements for the other five values of $\Gamma_x$ were conducted at 31.3$^{\circ}$C, corresponding to $Pr=5.25$. Previous studies have revealed that with increasing Prandtl number $Nu$ first increases, reaches its maximum value at around $Pr\approx4$ (depends on $Ra$), and then decreases slightly or remains independent of $Pr$ \cite[]{ahlers2001prl, gl2001prl, xia2002prl, ssv2010jfm, stevens2011jfm}. Therefore, within the present $Pr$ range, $Nu$ is expected to depend very weakly on $Pr$. Indeed, as shown in figures \ref{fig:fig3} and \ref{fig:fig5}, no significant differences are observed between the $Pr=5.45$ (circles) and 7.01 (right-triangles) results. Another source of uncertainty in the measured $Nu$ could be the non-Boussinesq effects, as some of our measurements were made at larger $\Delta$. \cite{funfschilling2005jfm} argued that the applied temperature difference $\Delta$ should be limited to $\lesssim15^{\circ}$C to strictly conform to the Boussinesq conditions. However, it can be seen from figure \ref{fig:fig2} and tables \ref{tab:tab1} and \ref{tab:tab2} that some of our data have $\Delta$ much larger than $15^{\circ}$C. As we shall see below, the measured large-$\Delta$ data show the same trend as those of small $\Delta$. This suggests that some of our data being not strictly Boussinesq will not change the main conclusions of the present work. Indeed, \cite{ahlers2006jfm} have shown that for water as the working fluid the non-Boussinesq effects could only slightly reduce the measured $Nu$ by at most $1.4\%$ for $\Delta=40$ K and thus $Nu$ is rather insensitive against even significant deviations from the Boussinesq conditions. Hence, for completeness all data are listed in the tables and are plotted in the figures.

Figure \ref{fig:fig3} shows $Nu/Ra^{1/3}$ as a function of $Ra$ for all six $\Gamma_x$. It is seen that data points for $\Gamma_x=20.8$ (down-triangles) collapse well on top of those for $\Gamma_x=14.3$ (diamonds) (within their overlap $Ra-$range), which in turn collapse well on top of those for $\Gamma_x=7.3$ (up-triangles). This implies that all sets of data can be described by a single curve over such a wide range of $\Gamma_x$, i.e., no significant $\Gamma_x$-dependence is observed. As discussed in $\S$1, $\Gamma$-dependence of $Nu$ essentially reflects the influence of flow structures on heat transport characteristics. Indeed, using particle image velocimetry (PIV), \cite{xia2008} have shown in rectangular cells that the number of the convection rolls depends systematically on the aspect ratio of the system: only one convection roll is observed in the $\Gamma_x=1$ and 2 cells and the LSC breaks into (horizontally arranging) multi-roll structure for $\Gamma_x$ larger than or equal to 4. Therefore, our present results suggest that for rectangular geometry turbulent heat transport is very insensitive to the nature and structures of the large-scale mean flows of the system. We note that our present results are different from those obtained in 2D numerical case \cite[]{lohse2011pre}, which were made for both $Pr=0.7$ and 4.3. For the 2D simulation, the stable states with $n$ rolls are found to enable larger heat transfer than those with $n+1$ rolls for vertically arranging LSC rolls. One possible reason for this difference may be attributed to different alignments of the LSC rolls, i.e. the aspect ratios of the 2D simulation vary between 0.4 and 1.25 and the LSC rolls are stacked vertically, whereas for the present geometry we have horizontally stacked rolls \cite[]{xia2008}.

%The reason for such a difference is not known at the present and deserves further studies.  with about $8\sim9\%$ difference in magnitude

\begin{figure}
\begin{center}
\resizebox{0.7\columnwidth}{!}{%
  \includegraphics{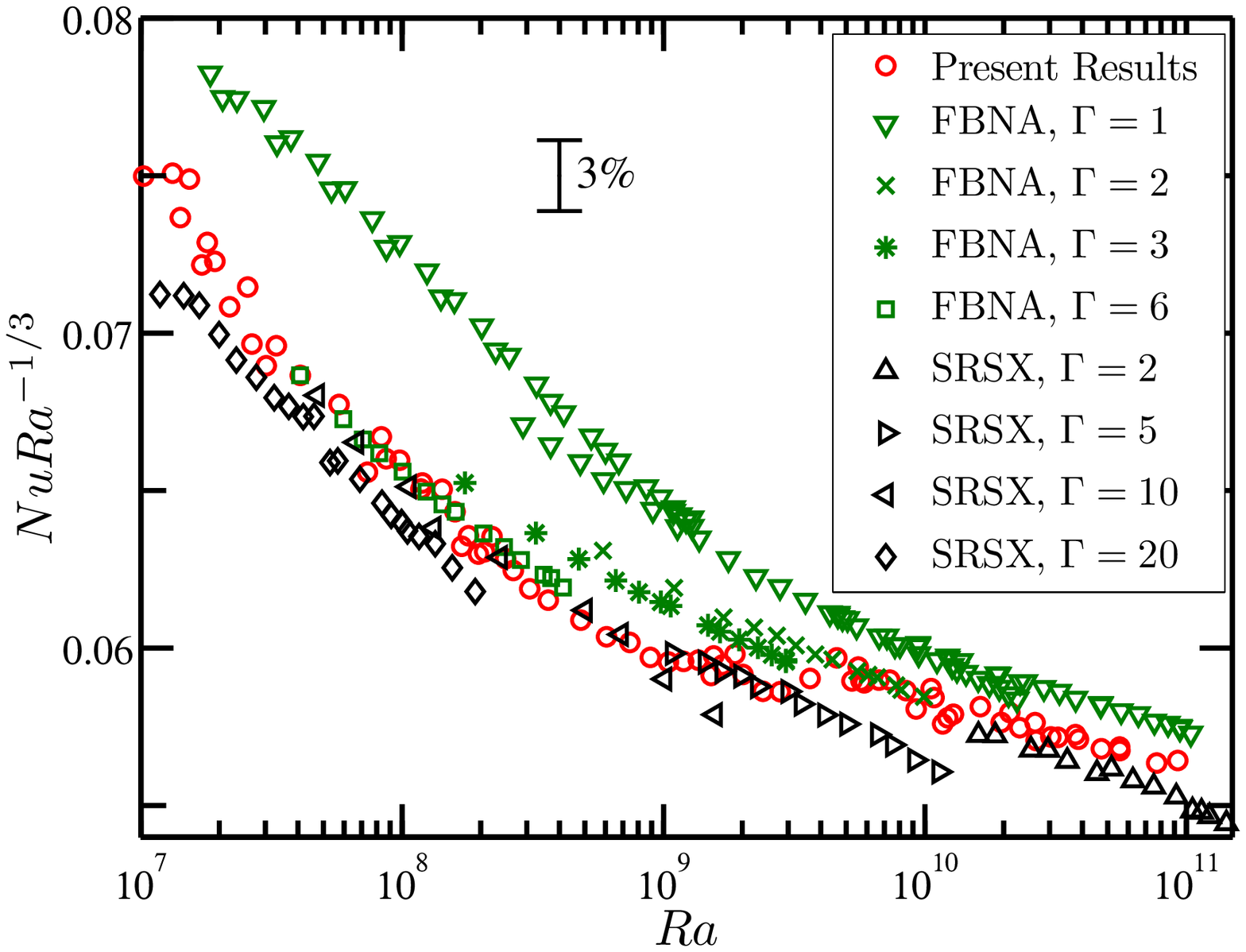}
}
%\resizebox{0.495\columnwidth}{!}{%
 % \includegraphics{fig4r}
%}
\caption{(color online). Comparison among $Nu/Ra^{-1/3}$ from the present work ($\bigcirc$), from \cite{funfschilling2005jfm} ($\bigtriangledown$, $\Gamma=1$; $\times$, $\Gamma=2$; $*$, $\Gamma=3$; $\square$, $\Gamma=6$; for clarity their $\Gamma=1.5$ data are not shown) and from \cite{sun2005jfm} ($\bigtriangleup$, $\Gamma=2$; $\vartriangleright$, $\Gamma=5$; $\vartriangleleft$, $\Gamma=10$; $\bigcirc$, $\Gamma=20$). As our results show that for rectangular geometry $Nu/Ra^{-1/3}$ is independent of the cells' aspect ratio, here we do not distinguish our data for different $\Gamma_x$. Note that all data are as measured, without the correction for the finite plate conductivity.} \label{fig:fig4}
\end{center}
\end{figure}

\begin{figure}
\begin{center}
\resizebox{0.495\columnwidth}{!}{%}
  \includegraphics{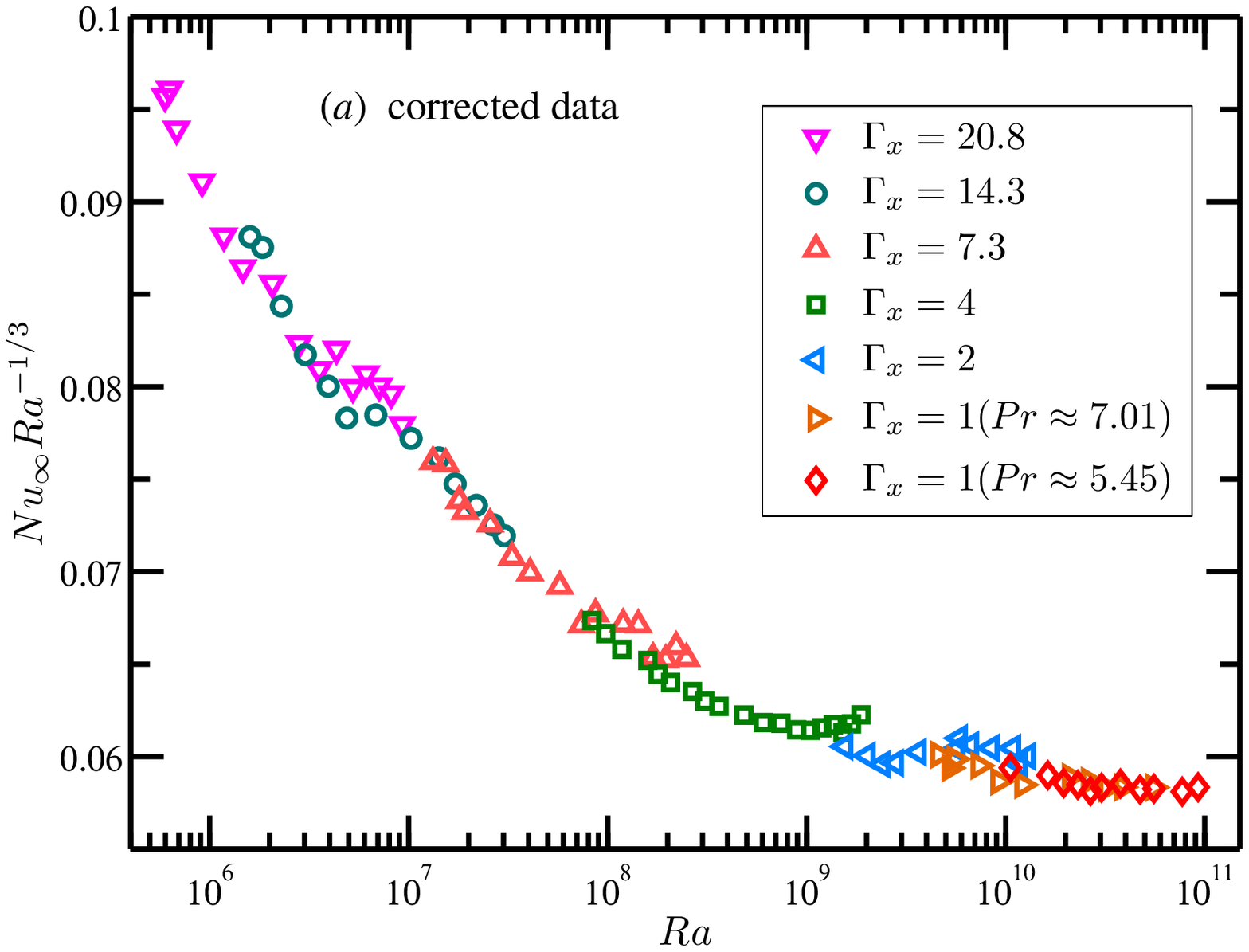}
}
\resizebox{0.495\columnwidth}{!}{%}
  \includegraphics{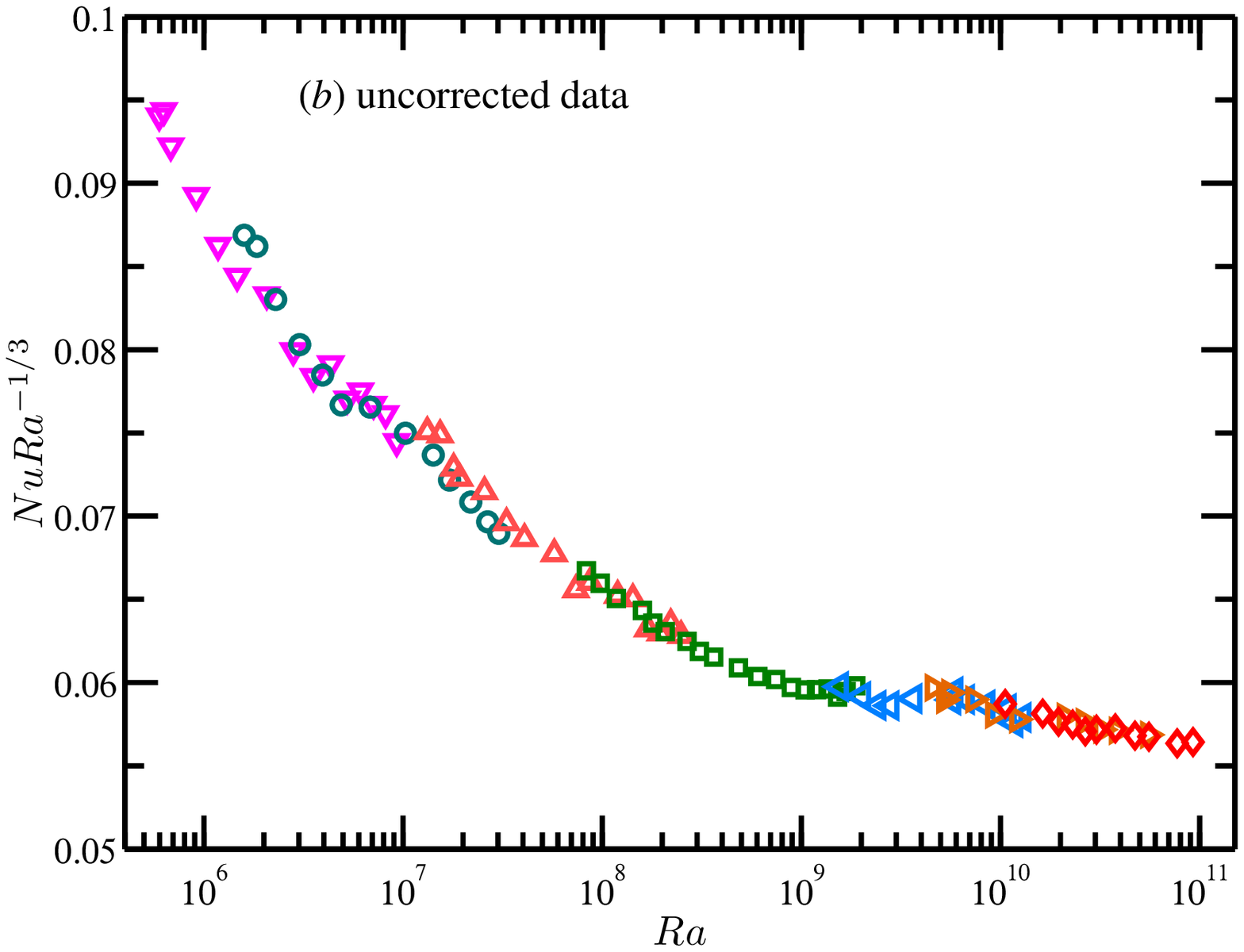}
}
\caption{(color online). (\emph{a}) Compensated $Nu_{\infty}/Ra^{1/3}$ on a linear scale vs. $Ra$ on a logarithmic scale: symbols as figure \ref{fig:fig3}. Here, $Nu_{\infty}$ is calculated using Eq. (\ref{eq:1}) and the parameters $a=0.275$ and $b=0.39$ obtained by \cite{brown2005pof} in cylindrical cells. For comparison, the uncorrected data $Nu/Ra^{1/3}$ of figure \ref{fig:fig3} are replotted in (\emph{b}).} \label{fig:fig5}
\end{center}
\end{figure}

In figure \ref{fig:fig4}, we compare the present results with those obtained in cylindrical cells and at $Pr\approx4.4$. We mainly consider two recent data sets: one is from \cite{funfschilling2005jfm} (referred to as FBNA) and the other is from \cite{sun2005jfm} (referred to as SRSX). As our measured $Nu$ is independent of the cells' aspect ratio, here we do not distinguish our data for different $\Gamma_x$.  Note that all data in figure 4 are as measured, without the correction for the finite plate conductivity. For the FBNA data, one sees that the $\Gamma=6$ data of FBNA are in excellent agreement with ours, but their small-$\Gamma$ data are a few percent larger. Nevertheless, we note that the $\Gamma=1$ data of FBNA (down-triangles) show similar trends to ours for $Ra\lesssim10^9$ and $Ra\gtrsim10^{10}$ and thus the small difference between the two sets of data may be attributed to different system errors. The SRSX data and ours are much closer. One sees that parts of the $\Gamma=5$ and 10 data of SRSX agree well with ours and the others lie slightly below our measured $Nu$.

%with about $2\sim3\%$ deviations.Such large deviations, especially for small $Ra$, could not be attributed to different $Pr$ of the FBNA and the present experiments. As reported by \cite{nikolaenko2005jfm}, over the ranges $2\times10^{10}\leq Ra\leq4\times10^{11}$ and $3.62\leq Pr\leq5.42$, $Nu$ scales as $Nu\sim Pr^{-0.044}$. This weak $Pr$-dependence could only introduce $0.8\%$ difference between $Pr=4.38$ (FBNA) and 5.25 (present) and $2\%$ difference between $Pr=4.38$ (FBNA) and 7.01 (present), which are both smaller than the observed deviations.

One noticeable difference between the present results and those of FBNA and SRSX is worthy of note. For both FBNA and SRSX the larger-$\Gamma$ results lie consistently below those of smaller ones, i.e. $Nu$ is generally smaller for larger $\Gamma$. This is true even for the largest-$\Gamma$ (i.e. $\Gamma=10$ and 20) data of SRSX (see figure \ref{fig:fig4}). Whereas, our measured $Nu$ for all six values of $\Gamma_x$ fall into a single curve, i.e. $Nu$ is essentially independent of $\Gamma_x$ for our results. To understand such a difference, we note that all data plotted in figure \ref{fig:fig4} have not been corrected for the influence of the finite conductivity of the top and bottom plates \cite[]{verzicco2004pof}. In a cylindrical cell of $\Gamma=0.5$, Sun, Xi $\&$ Xia (2005\emph{b}) argued that because of the finite conductivity and finite heat capacity of the plates the azimuthal sweeping of the circulation plane of the LSC would make heat transfer more efficient than the case when the LSC is locked in a particular orientation. Moreover, \cite{xi2008pre} studied the azimuthal motion of the LSC in cylindrical cells of $\Gamma=0.5$, 1.0, and 2.3 and their results showed that the LSC's azimuthal motion is more confined in larger-$\Gamma$ cells. If we generalize the above two findings to large $\Gamma$ and taken them together, a possible scenario can be achieved: the larger-$\Gamma$ cell confines the azimuthal sweeping motion of the LSC which in turn reduces the measured $Nu$. This scenario could be valid for cylindrical cells due to their azimuthal symmetry. But for rectangular cells, this is not expected to work because the rectangular geometry has already locked the orientation of the LSC \cite[]{xia2003pre}. Therefore, our present $\Gamma_x$-independent results are consistent with the lack of the azimuthal sweeping motion of the LSC in rectangular cells. What we should stress is that the large-$\Gamma$ cell contains multi-roll structures of the LSC \cite[]{thess2007pre,xia2008,schumacher2010jfm} and thus the azimuthal motion of the LSC and its influence on heat transfer should be more complicated for large $\Gamma$. However, as we shall see in figure \ref{fig:fig7}, when corrections for the finite conductivity of the plates are made, $\Gamma$-dependence of $Nu_{\infty}$ becomes weaker for both the FBNA and SRSX data, which suggests that the finite plate conductivity effect is indeed a major factor for the observed difference of the behaviors of heat transfer in rectangular and cylindrical cells.

\subsection{$Nu_{\infty}$ vs. $Ra$ for different $\Gamma_x$}

\cite{brown2005pof} suggested an empirical correction factor $f(X)=1-exp[-(aX)^b]$, namely
\begin{equation}
Nu=f(X)Nu_{\infty}=\{1-exp[-(aX)^b]\}Nu_{\infty}(Ra,Pr),
\label{eq:1}
\end{equation}
to obtain estimates of the ideal Nusselt number $Nu_{\infty}$ for plates with perfect conductivity from the measured $Nu$. Here, $X=R_f/R_p$ is the ratio of the effective thermal resistance of the working fluid, $R_f=H/(\lambda_fNu)$, to the thermal resistance of the plates, $R_p=e/\lambda_p$, $\lambda_p$ ($= 401$ W/m K) is the conductivity of plates (Cu), $\lambda_f$ ($\lambda_f=0.614$ W/m K for $Pr=5.25$ and $\lambda_f=0.589$ W/m K for $Pr=7$) is the conductivity of water, and $e$ ($=1.5$ cm) is the mean thickness of the top (the part of the top plate below the cooling channels is used here) and bottom plates \cite[]{verzicco2004pof}. To apply the relation (\ref{eq:1}) to our measured $Nu$, one needs to determine the values of $a$ and $b$. The best way to do this is to use plates of different thermal conductivities, as was done by \cite{brown2005pof}, who used two sets of plates made of Cu and Al. However, the lack of aluminum-plate measurements in the present study prevents us to determine the values of $a$ and $b$. Alternatively, to estimate $Nu_{\infty}$, we use the parameters $a=0.275$ and $b=0.39$ obtained by \cite{brown2005pof} in cylindrical samples of 50 cm in diameter.

\begin{figure}
\begin{center}
\resizebox{0.7\columnwidth}{!}{%
  \includegraphics{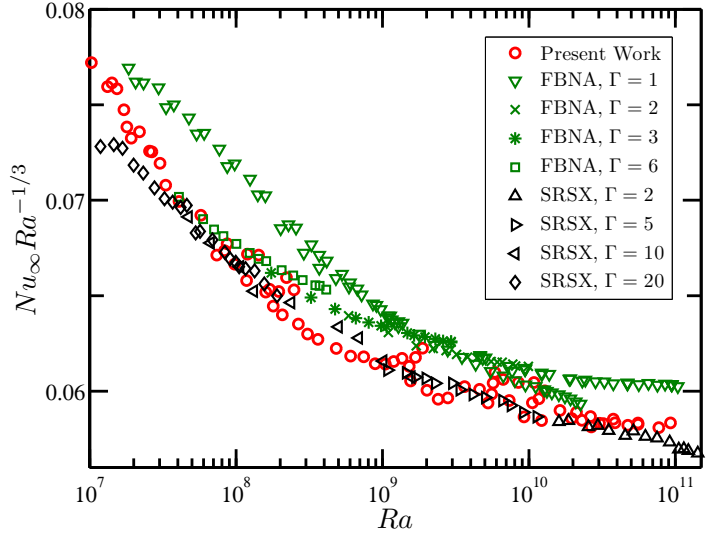}
}
\caption{(color online). Comparison among $Nu_{\infty}/Ra^{-1/3}$ from the present work, from \cite{funfschilling2005jfm}, and from \cite{sun2005jfm}. Note that the data have been corrected for the finite plate conductivity.} \label{fig:fig7}
\end{center}
\end{figure}

Figure \ref{fig:fig5}(\emph{a}) shows the calculated $Nu_{\infty}/Ra^{1/3}$ as a function of $Ra$. For comparison, the uncorrected data $Nu/Ra^{1/3}$ of figure \ref{fig:fig3} are replotted in figure \ref{fig:fig5}(\emph{b}). The corrected data in figure \ref{fig:fig5}(\emph{a}) seem to display a small aspect ratio dependence, e.g., near $Ra=2\times10^8$ the $\Gamma_x=7.3$ data lie slightly above the $\Gamma_x=4$ data and near $Ra=2\times10^9$ the $\Gamma_x=4$ data lie slightly above the $\Gamma_x=2$ data. We note that this dependence is consistent with the FBNA data (see figure \ref{fig:fig7}, where data points for $\Gamma=6$ lie slightly above those for $\Gamma=3$, which in turn lie slightly above those for $\Gamma=2$). Nevertheless, the differences of $Nu_{\infty}$ for the two adjacent-$\Gamma_x$ sets of data are only a few percent, which is smaller than or comparable with the experimental errors. Hence, we may also conclude that no significant $\Gamma_x$-dependence of heat transfer is observed.

We want to emphasize that here we adopted the parameters of $a=0.275$ and $b=0.39$ determined by \cite{brown2005pof} for cylindrical samples with a diameter of 50 cm. \cite{brown2005pof} have shown experimentally that $a$ and $b$ are independent of the aspect ratio, but vary with the diameter of the sample. In the present study, we chose the rectangle as the geometry of the convection cells, which can possibly have an influence on the finite plate correction, as we have discussed at the end of $\S$ 3.1. However, without the measurement performed with aluminum plates it is difficult to assess whether the geometry would (significantly) influence the values of $a$ and $b$. By using different sets of $a$ and $b$ to perform the finite conductivity correction, we estimate that the uncertainties of $a$ and $b$ may yield a few percent uncertainty on the obtained $Nu_{\infty}$, which is the same order of the difference between $Nu$ and $Nu_{\infty}$. We should also stress that the best way to estimate $Nu_{\infty}$ is to use plates with different thermal conductivities, as was done by \cite{brown2005pof} for cylindrical samples. Therefore, new measurements with Al plates are essential for concretely settling the problem and this will be the objective of future studies.

In figure \ref{fig:fig7} we directly compare our $Nu_{\infty}$ and those of FBNA and SRSX. It is seen that our data display the similar $Ra$-dependence as the $\Gamma=1$ data of FBNA, and an excellent agreement between our data and those of SRSX can be found for nearly all the overlap $Ra$ range. As the three sets of data were taken independently from the samples with very different geometries, this agreement is just remarkable.

\begin{table}
\centering \caption{Fitted parameters from equations (\ref{eq:Nu_single}).} %and (\ref{eq:Nu_inf}).}
\label{tab:tab3}
\begin{tabular}{c c c c c c c c c c c c c}
$\Gamma_x$ & & 1 & & 2 & & 4 & & 7.3 & & 14.3 & & 20.8\\[3pt]
$A_(\Gamma_x)$ & & 0.074 & & 0.059 & & 0.108 & & 0.173 & & 0.212 & & 0.242\\
$\beta(\Gamma_x)$ & & 0.324 & & 0.335 & & 0.307 & & 0.282 & & 0.271 & & 0.262\\
\end{tabular}
\end{table}

\begin{figure}
\begin{center}
%\resizebox{0.6\columnwidth}{!}{%
%  \includegraphics{fig8a}
%}
\resizebox{0.6\columnwidth}{!}{%
  \includegraphics{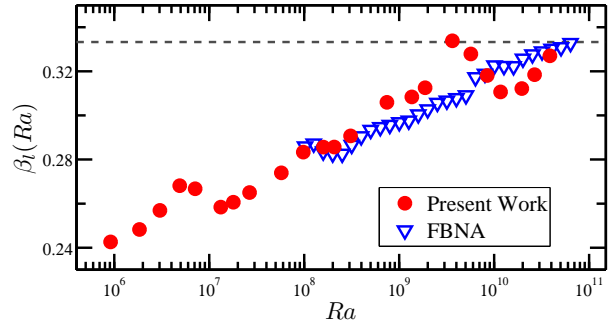}
}
\caption{(color online). Local scaling exponent $\beta_l$ of $Nu_{\infty}(Ra)$, determined from a power-law fit over a sliding window of half a decade, as a function of $Ra$ for the present data (solid circles) and for the $\Gamma=1$ data of \cite{funfschilling2005jfm} (open triangles). Dashed line marks $\beta_l=1/3$ for reference. Note that both sets of $\beta_l$ were obtained from the corrected data.} \label{fig:fig8}
\end{center}
\end{figure}

Finally, we studied the power-law relation between $Nu_{\infty}$ and $Ra$, namely,
\begin{equation}
Nu_{\infty}=A(\Gamma_x)Ra^{\beta(\Gamma_x)}.
\label{eq:Nu_single}
\end{equation}
Table \ref{tab:tab3} displays the fitted results for the power-law relation (\ref{eq:Nu_single}) at fixed aspect ratios. One sees that as $\Gamma_x$ increases in general the prefactor $A$ increases and the scaling exponent $\beta$ decreases. Here, $\Gamma_x$-dependences of $A$ and $\beta$ essentially reflect their $Ra$-dependence as our results do not reveal significant aspect-ratio dependence of $Nu_{\infty}$. We thus do not distinguish the data for different $\Gamma_x$ and take them together to study the scaling behaviors of $Nu_{\infty}(Ra)$. The local scaling exponent $\beta_l$ is obtained by a power-law fit, $Nu_{\infty}\sim Ra^{\beta_l}$, to the data for $Nu_{\infty}(Ra)$ within a sliding window that covers half a decade of $Ra$. Figure \ref{fig:fig8} shows the results for $\beta_l$ as a function of $Ra$. It is seen that $\beta_l$ roughly increases linearly with $\log{Ra}$ from $\beta_l=0.243$ at $Ra\approx9\times10^5$ to $\beta_l=0.342$ at $Ra\approx3.6\times10^9$. At small $Ra$, our results differ very much from the DNS results for cylindrical samples of unit aspect ratio by Wagner, Shishkina $\&$ Wagner (2012), who found that $\beta_l$ increases again as $Ra$ decreases below $2\times10^7$ and grows to about 0.30 at $Ra=10^6$ [see figure 2 of Wagner \emph{et al.} (2012)]. The apparent differences in $\beta_l$ may be explained by the different Prandtl number in the two studies, i.e., Wagner \emph{et al.} (2012) carried out their simulations at $Pr=0.786$, while our measurements were made at $Pr\approx5.45$. For higher $Ra$, $\beta_l$ drops slightly and fluctuates around 0.32. The FBNA results for $\beta_l$ are also displayed together with ours in figure \ref{fig:fig8} for comparison. For $Ra\geq10^8$ the two data sets both increase with $Ra$ with ours being a little scatter. A source of uncertainty for our data could be the non-Boussinesq effect, i.e. the FBNA data were obtained in the strictly Boussinesq range, while some of our data are beyond the Boussinesq range (i.e. $\Delta>15^{\circ}$C). However, as have discussed in $\S$3.1, some of our data being not strictly Boussinesq will not change our main conclusions. What is worthy of note is that around $Ra\approx10^{10}$ our measured $\beta_l$ has a value that is close to the value of 1/3. The exponent $\beta\approx1/3$ was obtained before in cylindrical cells of $\Gamma=1$ by FBNA at $Ra\approx7\times10^{10}$ (see open triangles in figure \ref{fig:fig8}) and of $\Gamma=4$ by \cite{niemela2006jfm} for $Ra>10^{10}$. Here, our results in rectangular cells seem to be qualitatively consistent with these findings

\section{Conclusion}
In conclusion, our high-precision measurements of $Nu$ in rectangular cells with $(\Gamma_x, \Gamma_y)$ varying from $(1,0.3)$ to $(20.8, 6.3)$ show that $Nu$ is independent of the aspect ratio. This is slightly different from the observations by both FBNA and SRSX in cylindrical cells where $Nu$ is found to be in general a decreasing function of $\Gamma$, at least for $\Gamma\sim 1$ and larger. Such a difference may be attributed to different azimuthal dynamics of the large-scale circulation (LSC) and is probably a manifestation of the finite plate conductivity effect. To make finite conductivity corrections, an empirical correction factor $f(X)=1-exp[-(aX)^b]$, together with the parameters $a=0.275$ and $b=0.39$ obtained by \cite{brown2005pof} in cylindrical samples, were adopted to estimate $Nu_{\infty}$ for plates with perfect conductivity from the measured $Nu$. The obtaiend $Nu_{\infty}$ were found to be consistent with the FBNA and SRSX data measured in cylindrical samples to only a few percent. The scaling behaviors between $Nu_{\infty}$ and $Ra$ were studied for all six aspect ratios. The local scaling exponents $\beta_l$ of $Nu_{\infty}\sim Ra^{\beta_l}$ were calculated and found to increase with increasing $Ra$. Around $Ra\approx10^{10}$ our measured $\beta_l$ has a value that is close to the value of $1/3$.

\begin{acknowledgments}
This work was supported by the Natural Science Foundation of China (Nos. 11161160554, 11002085, and 11032007) and Shanghai Program for Innovative Research Team in Universities.
\end{acknowledgments}
%\bibliographystyle{jfm}
%\bibliography{all}

\end{document}